\documentclass[twoside,11pt]{article}

\usepackage{obs_study_style}
\usepackage{amsmath}
\usepackage{booktabs}

\usepackage{xcolor}
\definecolor{qz}{RGB}{255,0,0}

\newcommand{\pkg}[1]{{\normalfont\fontseries{b}\selectfont #1}}
\newcommand{\reals}{\mathbb{R}}
\newcommand{\E}{\mathsf{E}}
\newcommand{\VAR}{\mathsf{VAR}}
\newcommand{\COV}{\mathsf{COV}}

\newcommand{\brct}[1]{\left(  #1  \right)}
\newcommand{\bma}[1]{\begin{pmatrix} #1 \end{pmatrix}}
\newcommand\code{\bgroup\@makeother\_\@makeother\~\@makeother\$\@codex}

\heading{}{}{}{}{}{Author Names}

\ShortHeadings{ivmodel for Instrumental Variables Analysis}{}
\firstpageno{1}

\begin{document}

\title{ivmodel: An R Package for Inference and Sensitivity Analysis of Instrumental Variables Models with One Endogenous Variable}

\author{\name Hyunseung Kang  \email hyunseung@stat.wisc.edu \\
       \addr Department of Statistics\\
       University of Wisconsin-Madison\\
       Madison, WI 53706, USA
       \AND
       \name Yang Jiang \email yang.jiang@uber.com \\
       \addr Uber Technologies, Inc. \\
       New York, NY, USA
       \AND
       \name Qingyuan Zhao \email qyzhao@statslab.cam.ac.uk \\
       \addr Statistical Laboratory \\
       University of Cambridge \\
       Cambridge, United Kingdom
       \AND
       \name Dylan S. Small \email dsmall@wharton.upenn.edu \\
       \addr Department of Statistics \\
       The Wharton School, University of Pennsylvania \\
       Philadelphia, PA 19104}

\maketitle

\begin{abstract}
   We present a comprehensive R software \textbf{ivmodel} for analyzing instrumental variables with one endogenous variable. The package implements a general class of estimators called $k$-class estimators and two confidence intervals that are fully robust to weak instruments. The package also provides power formulas for various test statistics in instrumental variables. Finally, the package contains methods for sensitivity analysis to examine the sensitivity of the inference to instrumental variables assumptions. We demonstrate the software on the data set from \citet{card_using_1995}, looking at the causal effect of levels of education on log earnings where the instrument is proximity to a four-year college.

\end{abstract}


\section{Introduction}
The instrumental variables (IV) method is a popular method to estimate the casual effect of a treatment, exposure, or policy on an outcome when there is concern about unmeasured confounding \citep{angrist_instrumental_2001, hernan_instruments_2006, baiocchi_instrumental_2014}. IV methods have been widely used in statistics \citep{angrist_identification_1996}, economics \citep{angrist_instrumental_2001}, genomics and epidemiology \citep{davey_smith_mendelian_2003}, sociology \cite{bollen_instrumental_2012}, psychology \citep{gennetian_statistical_2008}, political science \citep{sovey_instrumental_2011}, and countless other fields. We also note that instrumental variables have been used to correct for measurement errors; see \citet{fuller_measurement_2006} for a full treatment on measurement errors.

Informally speaking, IV methods rely on having variables called instruments which are related to the exposure and are exogenous. An instrument is exogenous if it only affects the outcome through the pathway of affecting the exposure (i.e. the instrument has no direct effect on the outcome) and is independent of unmeasured confounders; see Section \ref{sec:ivassumptions} for details. Typically, instruments either come from (i) natural experiments whereby the instruments are naturally assigned to individuals at random or (ii) an actual randomized experiment whereby the treatment randomization is used as an instrument. For example, in Mendelian randomization, natural genetic variations that occur at conception have been used as instruments to answer causal questions in epidemiology; usually the instruments are single nucleotide polymorphisms (SNPs) at a specific location in the human genome \citep{davey_smith_mendelian_2003, davey_smith_mendelian_2004, lawlor_mendelian_2008}. In \citet{sexton_clinical_1984} and \citet{permutt_simultaneous_1989}, the authors studied the effect of maternal smoking on birth weight by randomly assigning the healthcare provider of pregnant mothers to two different group, the first group where the provider was asked by the investigators to encourage mothers to stop smoking and the second group where the provider did not receive this request from the investigators. Table~\ref{tab:ivexamples} illustrates other examples of instrumental variables; for more examples, see \citet{angrist_instrumental_2001} and \citet{baiocchi_instrumental_2014}.

\begin{table}[h!]
\centering
\begin{tabular}{p{2.5cm} p{3.1cm} p{4.0cm} p{3.7cm}} \hline
Outcome & Exposure & Instruments & Reference \\ \hline
& & &\\
\multicolumn{4}{c}{Natural experiments / Mendelian randomization} \\ \hline
Earnings & Years of schooling & Proximity to college when growing up & \citet{card_using_1995} \\
Earnings & Years of schooling & Quarter of birth & \citet{angrist_does_1991} \\
Metabolic phenotypes & C-reactive protein (CRP) & SNPs rs1800947, rs1130864, rs1205 & \citet{timpson_c-reactive_2005} \\
Blood pressure & Alcohol intake & Alcohol dehydrogenase (ALDH2) genotype & \citet{chen_alcohol_2008} \\
 & & &\\
\multicolumn{4}{c}{Randomized experiments / Encouragement designs} \\ \hline
Birth weight & Mother's smoking & Randomized encouragement to stop smoking & \citet{sexton_clinical_1984} and \citet{permutt_simultaneous_1989}  \\
Test scores & Class size & Randomized assignment to different class sizes & \citet{krueger_experimental_1999}
\end{tabular}
\caption{Application of instrumental variables methods based on source of instruments. Natural experiments/Mendelian randomization refer to instrumental variables studies where the instruments come from natural sources, such as genes or calendar years. Randomized experiments/encouragement designs refer to instrumental variables studies where the instruments are based on actual randomization mechanisms.}
\label{tab:ivexamples}
\end{table}

Software for running instrumental variables methods varies widely depending on the programming language. For example, in STATA, there are comprehensive and unified programs to handle the most popular instrumental variables methods, most notably \textbf{ivreg2} \citep{baum_instrumental_2003, baum_instrumental_2007} and \textbf{ivregress}. In R, different types of instrumental variables methods are implemented in different packages, for instance \textbf{AER} by \citet{kleiber_applied_2008}, \textbf{sem} by \citet{fox_sem_2014}, and \textbf{lfe} by \citet{gaure_lfe_2013}. Unfortunately, these packages do not include (i) modern instrumental variables methods, especially confidence interval procedures, that are robust to weak instruments, (ii) power calculations for IV analysis, and (iii) sensitivity analysis methods that examine sensitivity of inference to violations of IV assumptions.

The goal of the paper is to present an R package \textbf{ivmodel} that conducts a comprehensive instrumental variables analysis when there is one exposure/endogenous variable. These functions include a general class of estimators known as $k$-class estimators; see Section \ref{sec:kClass} for details. The functions also include two methods for confidence intervals that are fully robust to weak instruments \citep{stock_survey_2002}, the Anderson and Rubin confidence interval \citep{anderson_estimation_1949} and the conditional likelihood ratio confidence interval \citep{moreira_conditional_2003}. The package includes functions to calculate power of tests. Finally, the package includes methods to conduct sensitivity analysis in order to examine the sensitivity of the IV analysis to violations of IV assumptions.

\section{Instrumental variables model for one endogenous variable}
\subsection{Notation} \label{sec:notation}


Let there be $n$ individuals indexed by $i=1,\ldots,n$. For each individual $i$, we observe outcome $Y_i \in \reals$, exposure $D_i \in \reals$, $L$ instruments $Z_{i \cdot} \in \reals^L$, and $p$ covariates $X_{i \cdot} \in \reals^p$. Let $Y = (Y_1,\ldots,Y_n) \in \reals^n$ denote the vector of outcomes, $D = (D_1,\ldots,D_n) \in \reals^n$ denote the vector of exposures,  $Z \in \reals^{n \times L}$ denote the matrix of instruments where the $i$th row corresponds to $Z_{i \cdot}$, and $X \in \reals^{n \times p}$ denote the matrix of covariates where the $i$th row corresponds to $X_{i \cdot}$. Let $W = [Z : X]$ where $W$ is an $n$ by $L + p$ matrix that concatenates the matrices $Z$ and $X$.

For any matrix $M$, denote its transpose as $M^\intercal$. Also, for any matrix $M$, let $P_{M} = M (M^\intercal M)^{-1}M^\intercal$ be the orthogonal projection matrix onto the column space of $M$ and $R_{M}$ be the residual projection matrix so that $R_{M} + P_{M} = I$ and  $I$ is an $n$ by $n$ identity matrix. We assume that $(M^T M)^{-1}$ is well-defined and has a proper inverse. Finally, for any vector $v \in \reals^n$, let ${\rm diag}(v)$ be the $n$ by $n$ diagonal matrix whose diagonal elements consist of $v_{1},\ldots,v_n$.

\subsection{Model}
\label{sec:model}
We assume the following linear structural model between the observed quantities, $Y_i, D_i, Z_{i \cdot}$, and $X_{i \cdot}$

\begin{equation} \label{eq:model}
Y_i = D_i \beta + X_{i \cdot}^T \kappa + \epsilon_i, \quad{} \E(\epsilon_i | Z_{i \cdot}, X_{i \cdot}) = 0, \VAR(\epsilon_i | Z_{i \cdot}, X_{i \cdot}) = \sigma^2
\end{equation}
This is the standard, single equation homoskedastic linear structural model in econometrics  \citep{wooldridge_econometrics_2010}; Section \ref{sec:hetero} discusses the heteroskedastic and clustered variance models where $\sigma^2$ may vary for each individual. Model \eqref{eq:model} is not the usual regression model because $D_i$ is potentially correlated with $\epsilon_i$.  The parameter of interest is $\beta$, which can be interpreted as the causal effect of the exposure $D_i$ on the outcome $Y_i$; see next paragraph for more details. The parameter $\kappa$ relates the $p$ covariates to the outcome. We remark that $X_{i \cdot}$ can contain a value of $1$ to represent the intercept.

The parameters in model \eqref{eq:model} can be given a causal interpretation by using the potential outcomes notation \citep{rubin_estimating_1974} and the additive, linear constant effects (ALICE) model \citep{holland_causal_1988}. Let $Y_i^{(d,z)}$ be the potential outcome if individual $i$ were to have exposure $d$, a scalar value, and $L$ instruments $z$. Let $D_i^{(z)}$ be the potential exposure if individual $i$ had $L$ instruments $z$. For each individual, only one realization of $Y_i^{(d,z)}$ and $D_{i}^{(z)}$ is observed, denoted as $Y_i$ and $D_i$, respectively, based on individual $i$'s observed instrument values $Z_{i \cdot}$ and exposure $D_{i}$. Then, for two possible values of the exposure $d',d$ and instruments $z',z$, we assume the following potential outcomes model
\begin{equation} \label{eq:modelPotentialOutcome}
Y_{i}^{(d',z')} - Y_{i}^{(d,z)} = (d' - d) \beta, \quad{} \E(Y_i^{(0,0)} \mid Z_{i \cdot}, X_{i \cdot}) =  X_{i \cdot}^T \kappa
\end{equation}
In model \eqref{eq:modelPotentialOutcome}, $\beta$ represents the causal effect (divided by $d' - d$) of changing the exposure from $d'$ to $d$ on the outcome. The parameter $\kappa$ represents the impact of covariates on the baseline potential outcome $Y_{i}^{(0,0)}$. If we further define $\epsilon_i = Y_{i}^{(0,0)} - \E(Y_{i}^{(0,0)} \mid Z_{i \cdot}, X_{i \cdot})$, we obtain the observed data model in \eqref{eq:model}, thus providing the parameters in the observed model in \eqref{eq:model} a causal interpretation.


We'll also introduce a model for the relationship between the endogenous variable $D_i$, the instruments $Z_{i \cdot}$, and the covariates $X_{i \cdot}$
\begin{equation} \label{eq:model_1st_stage}
D_i = Z_{i \cdot}^T \gamma + X_{i \cdot}^T \tilde\kappa + \eta_i, \quad{} \E(\eta_i | Z_{i \cdot}, X_{i \cdot}) = 0, \VAR(\eta_i | Z_{i \cdot}, X_{i \cdot}) = \omega^2
\end{equation}
This ``first stage'' model in \eqref{eq:model_1st_stage} is not necessary for all our methods in the \textbf{ivmodel} package. In particular, the $k$-class estimators in Section \ref{sec:kClass} and the confidence interval for the Anderson and Rubin test in Section \ref{sec:fullrobustCI} are valid without the first stage modeling assumption in \eqref{eq:model_1st_stage}. However, other methods presented in the paper require this model. Also, it's common in econometrics to assume a linear relationship between $D$, $Z$ and $X$ \citep{wooldridge_econometrics_2010}.



We conclude by simplifying the models in equations \eqref{eq:model} and \eqref{eq:model_1st_stage} by projecting out the covariates $X$ using the Frisch-Waugh-Lovell Theorem \citep{davidson_estimation_1993}. Specifically, models \eqref{eq:model} and \eqref{eq:model_1st_stage} are equivalent to
\begin{align} \label{eq:model-simplified}
Y_i^* = D_i^* \beta + \epsilon_i^* \\
D_i^* = Z_{i \cdot}^* \gamma + \eta_i^* \label{eq:model-simplified_1st_stage}
\end{align}
where
\[
Y^* = R_X Y, \quad{} D^* = R_X D, \quad{} Z^* = R_X Z, \quad{} \epsilon^* = R_X \epsilon, \quad{} \eta^* = R_X \eta
\]
The superscripts $Y^{*}, D^{*}, Z^{*}$ represent the outcome, the exposure, and the instruments after controlling for the covariates $X$ using the residual orthogonal projection $R_{X}$ defined in Section \ref{sec:notation}. The equivalent models \eqref{eq:model-simplified} and \eqref{eq:model-simplified_1st_stage} allow us to concentrate on the target parameter of interest, $\beta$, and simplify the expressions of the instrumental variables methods presented in the paper. 

\subsection{Assumption of instrumental variables} \label{sec:ivassumptions}
Under the model in \eqref{eq:model}, we make the standard assumptions in the instrumental variables literature below \citep{wooldridge_econometrics_2010}.
 \begin{itemize}
 \item[(A1)] $\E(W^T W)$ is full rank.
 \item[(A2)] Conditional on the covariates $X$, the instruments $Z$ are associated with the exposure $D$, $\E(Z^T R_X D) \neq 0$
 \item[(A3)] $W$ is exogenous, $\E(W^T \epsilon) = 0$
\end{itemize}
Assumption (A1) is a standard moment condition on the matrix of exogenous variables that include covariates and instruments. Assumption (A2) states that conditional on the covariates $X$, the instruments are associated with the exposure. There are many ways to test this assumption in practice, the most popular being the F statistic. Specifically, we would test whether the regression coefficients associated with $Z$ is zero in the regression of $D$ on $X$ and $Z$.  Instruments with $F$ statistics greater than $10$ are considered to be strong instruments while instruments with $F$ statistics below $10$ are considered to be weak instruments  \citep{stock_survey_2002}. Assumption (A3) is satisfied  in the ALICE model if $Z$ has no direct on $D$ and $Z$ is independent of unmeasured confounders. Assumption (A3) is generally untestable in that it's impossible to check whether the exogenous variables $Z$ and $X$ are uncorrelated with the structural error $\epsilon_i$, which is never observed. However, if there are more than one instruments, methods exist to partially test this assumption, the most popular being the Sargan's test \citep{sargan_estimation_1958}. Under all three assumptions (A1)-(A3), standard econometric arguments show that the the model parameters in \eqref{eq:model} are identified; see Section 5.2 of \citet{wooldridge_econometrics_2010}.

Typically, practitioners assume that instruments satisfy (A1)-(A3) and proceed with estimating the target parameter $\beta$ \citep{angrist_instrumental_2001}. However, violations of these assumptions occur, especially (A2) and (A3). If (A2) is weakly satisfied such that instruments $\E(Z^T R_X D) \approx 0$, also known as the weak instrument problem, the most commonly used instrumental variables estimation method, two stage least squares (TSLS), produces biased estimates of $\beta$ \citep{nelson_distribution_1990, staiger_instrumental_1997, stock_survey_2002}. Thankfully, many robust methods exist with weak instruments and we discuss them in Section \ref{sec:fullrobustCI}. Violation of (A3), known as the invalid instrument problem \citep{murray_avoiding_2006}, has received far less attention than the weak instrument problem, but some progress has been made in this area \citep{kolesar_identification_2015, kang_instrumental_2016, jiang_sensitivity_2017}. This paper presents one way to deal with violations of (A3) via a sensitivity analysis in Section \ref{sec:sensitivity}.

\section{$k$-class estimation and inference} \label{sec:kClass}
\subsection{Definitions and general properties} \label{sec:defkClass}
A class of estimators for $\beta$, called the $k$-class estimators and denoted as $\widehat{\beta}_{k}$, is defined as follows.
\begin{equation} \label{eq:kclassEst}
\widehat{\beta}_{k} = ({D^*}^T (I - k R_{Z^*}) D^*)^{-1} {D^*}^T (I - kR_{Z^*}) Y^*
\end{equation}
Table~\ref{tab:kclass} lists some estimators that are $k$-class estimators, including ordinary least squares (OLS), two-stage least squares (TSLS), limited information maximum likelihood (LIML), and Fuller's estimator (FULL). For example, the LIML estimator uses a $k_{LIML}$ which is the minimum value of $k$ that satisfies the following equation
\begin{equation}
det
\begin{pmatrix}
{Y^*}^T(I - k R_{Z^*}) Y^* & {Y^*}^T (I - k R_{Z^*}) D^* \\
{D^*}^T (I - k R_{Z^*}) Y^* & {D^*}^T (I - k R_{Z^*}) D^*
\end{pmatrix}
 = 0
\end{equation}

\begin{table}[h!]
\centering
\begin{tabular}{p{5.2cm} l }
$k$ & Name \\ \hline
$k = 0$ & Ordinary least squares (OLS) \\
$k = 1$ & Two-stage least squares (TSLS) \\
$k = k_{LIML}$ & Limited information maximum likelihood (LIML) \\
$k = k_{LIML} - \frac{b}{n - L - p}, b > 0$ & Fuller's estimator (FULL)
\end{tabular}
\caption{Different types of $k$-class estimator}
\label{tab:kclass}
\end{table}

Each $k$ yields an estimator with unique finite-sample properties, which will be discussed in detail in Section \ref{sec:kclassexamples}. But, asymptotically, all $k$-class estimators are consistent for $\beta$ when $k \to 1$ as $n \to \infty$\citep{davidson_estimation_1993}.
In addition, when $\sqrt{n}(k - 1) \to 0$ as $n \to \infty$, the $k$-class estimator has an asymptotic Normal distribution \citep{amemiya_advanced_1985}
\begin{equation} \label{eq:kclassasymptotic}
\frac{\widehat{\beta}_{k} - \beta}{\sqrt{\widehat{\VAR}(\widehat{\beta}_k)}} \to N(0,1)
\end{equation}
where
\begin{equation} \label{eq:stderror}
\widehat{\VAR}(\widehat{\beta}_k) = \widehat{\sigma}^2 ({D^*}^T(I - k R_{Z^*}) {D^*})^{-1}, \quad{} \widehat{\sigma}^2 = \frac{(Y^* - D^* \widehat{\beta}_{k})^T (Y^* - D^* \widehat{\beta}_{k})}{n - p-1}
\end{equation}
The asymptotic distribution in \eqref{eq:kclassasymptotic} allows us to test hypotheses
\begin{equation} \label{eq:hyp}
H_0: \beta = \beta_0, \quad{} H_a: \beta \neq \beta_0
\end{equation}
by comparing the standardized deviate in \eqref{eq:kclassasymptotic} to a standard Normal, or a $t$ distribution with degrees of freedom $n - L - p$. We can also create a $1 - \alpha$ confidence interval for $\beta$ with $\widehat{\beta}_{k}$, i.e.
\[
\left(\widehat{\beta}_{k} - z_{1-\alpha/2} \sqrt{\widehat{\VAR}(\widehat{\beta}_k)}, \quad{} \widehat{\beta}_{k} + z_{1-\alpha/2} \sqrt{\widehat{\VAR}(\widehat{\beta}_k)} \right)
\]
where $z_{1-\alpha/2}$ is the $1 -\alpha/2$ quantile of the standard Normal distribution. We can alternatively use the $1 -\alpha/2$ quantile of the $t$ distribution with  $n - L - p$ degrees of freedom.

\subsection{Some Examples of $k$-class Estimators} \label{sec:kclassexamples}
The most well-known $k$-class estimator in instrumental variables is two-stage least squares (TSLS) where $k = 1$, i.e.
\[
\widehat{\beta}_{1} = ({D^*}^T P_{Z^*} D^*)^{-1} {D^*}^T P_{Z^*} Y^*
\]
In addition to being consistent and having an asymptotic Normal distribution, TSLS is efficient among all IV estimators using linear combination of instruments $Z$ (Theorem 5.3 in \citet{wooldridge_econometrics_2010}). In fact, under the asymptotics rates of $\sqrt{n}(k - 1) \to 0$ discussed in the prior section, all $k$-class estimators have the same asymptotic Normal distribution as TSLS. Also, when $L = 1$, TSLS and LIML produce identical estimates of $\beta$ \citep{davidson_estimation_1993}.

Despite having the same asymptotic distribution, each $k$-class estimators behave differently in finite-samples. With weak instruments, TSLS tends to be biased towards OLS in finite sample and the bias may persist even with large samples \citep{bound_problems_1995}. In contrast, LIML and FULL are more robust to violations of (A2) than TSLS \citep{stock_survey_2002}. However, LIML has no finite moments while TSLS has up to $L - 1$ moments. FULL corrects LIML's lack of moments by having moments if the sample  is large enough \citep{davidson_estimation_1993}.

Other types of $k$-class estimators exist beyond those listed in Table \ref{tab:kclass} and no single $k$-class estimator uniformly dominates another in all settings \citep{davidson_estimation_1993}. In practice, the most popular estimators are TSLS and LIML, with LIML being more robust to weak instruments  \citep{stock_survey_2002, mariano_simultaneous_2003, chao_consistent_2005}

\subsection{Heteroskedasticity and Clustering when $k = 1$} \label{sec:hetero}
When model \eqref{eq:model} has heteroskedastic variance or cluster-level variance, a $k$-class estimator with $k = 1$ can be modified to obtain correct standard errors for the estimate $\widehat{\beta}_{1}$. Specifically, under heteroskedasticity where $\VAR(\epsilon_i \mid Z_{i \cdot}, X_{i \cdot}) = \sigma_i^2$, we would replace the estimator of $\widehat{\VAR}(\widehat{\beta}_k)$ in equation \eqref{eq:stderror} with the heteroskedastic-consistent estimator of variance proposed in \citet{white_heteroskedasticity_1980}.
\begin{equation} \label{eq:hetero_se}
\widehat{\VAR}_{HC}(\widehat{\beta}_1)  =  {D^*}^\intercal R_{Z^*} {\rm diag}(Y^* - D^* \widehat{\beta}_{1}) R_{Z^*} {D^*} ({D^*}^\intercal R_{Z^*} {D^*})^{-2}
\end{equation}
Under clustering where we have $C$ clusters, $\VAR(\epsilon_i \mid Z_{i \cdot}, X_{i \cdot}) = \sigma_j^2$ for each cluster $j \in \{1,\ldots,C\}$, and $\COV(\epsilon_i, \epsilon_{i'} \mid Z_{i \cdot}, X_{i \cdot}) = 0$, we can use the same variance estimator in equation \eqref{eq:hetero_se} \citep{cameron_practicioner_2015}.

\section{Dealing with weak instruments: Robust confidence intervals} \label{sec:fullrobustCI}

In this section, we discuss the case when the instruments may nearly violate (A2) and discuss two inferential procedures that are fully robust to violations of (A2).

Let $M$ be an $n$ by $2$ matrix where the first column contains $Y^*$ and the second column contains $D^*$. Let $a_0 = (\beta_0, 1)$ and $b_0 = (1, -\beta_0)$ to be two-dimensional vectors and $\widehat{\Sigma} = M^T R_{Z^*} M / (n - L - p)$.  Let $\widehat{S}$ and $\widehat{T}$ be two-dimensional vectors defined as follows.
\[
\widehat{S} = \frac{ ({Z^*}^T Z^*)^{-1/2} {Z^*}^T M b_0}{\sqrt{b_0^T \hat{\Sigma} b_0}}, \quad{}
\widehat{T} = \frac{ ({Z^*}^T Z^*)^{-1/2} {Z^*}^T M \widehat{\Sigma}^{-1} a_0}{\sqrt{a_0^T \widehat{\Sigma}^{-1} a_0}}
\]
We also define the following scalar values, $\widehat{Q}_1, \widehat{Q}_2$, and $\widehat{Q}_3$.
\[
\widehat{Q}_{1} = \widehat{S}^T \widehat{S}, \quad{} \widehat{Q}_{2} = \widehat{S}^T \widehat{T}, \quad{}\widehat{Q}_{3} = \widehat{T}^T \widehat{T}
\]
Based on $\widehat{Q}_{1}$, $\widehat{Q}_{2}$, and $\widehat{Q}_{3}$, we define two tests for testing $H_0: \beta = \beta_0$ that are fully robust to violations of (A2), the Anderson-Rubin test \citep{anderson_estimation_1949}, and the conditional likelihood test \citep{moreira_conditional_2003}.
\begin{align}
AR(\beta_0) &= \frac{\widehat{Q}_{1}}{L} \label{eq:AR} \\
CLR(\beta_0) &= \frac{1}{2} (\widehat{Q}_{1} - \widehat{Q}_{3}) + \frac{1}{2}\sqrt{(\widehat{Q}_{1} + \widehat{Q}_{3})^2 - 4(\widehat{Q}_{1} \widehat{Q}_{3} - \widehat{Q}_{2}^2)} \label{eq:CLR}
\end{align}
Many works have shown that these two tests are fully robust to weak instruments in that even if the instrument strength is near zero, the two tests still maintain Type I error control \citep{staiger_instrumental_1997, stock_survey_2002, moreira_conditional_2003, dufour_identification_2003, andrews_optimal_2006}. Between the two tests, there is no uniformly most powerful test under weak instruments, but \citet{andrews_optimal_2006} and \citet{mikusheva_robust_2010} suggest using \eqref{eq:CLR} due to its generally favorable power compared to \eqref{eq:AR} in most scenarios. However, the Anderson-Rubin test is the simplest of the two tests in that under a Normal error assumption, it can be written as a standard F-test in regression where the outcome is $R_{Z^*}(Y - D\beta_0)$, the regressors are $Z^*$, and we are testing whether the coefficients associated with the regressors $Z^*$ are zero or not using an F-test. Also, the Anderson-Rubin test in \eqref{eq:AR} does not require the first stage model in \eqref{eq:model_1st_stage} \citep{dufour_identification_2003} whereas the conditional likelihood ratio test does.

We can invert both tests in equation \eqref{eq:AR} and \eqref{eq:CLR} to obtain $1 -\alpha$ confidence intervals that are fully robust to weak instruments, i.e. $\{\beta : AR(\beta_0) \leq F_{L,n-L-p, 1 - \alpha}\}$ for the Anderson-Rubin confidence interval and $\{\beta : CLR(\beta_0) \leq q_{1-\alpha} \}$ for the conditional likelihood ratio test. Here, $F_{L,n-L -p,1-\alpha}$ is the $1-\alpha$ quantile of the $F$ distribution with  $L$ and $n - L - p$ degrees of freedom. The term $q_{1-\alpha}$ is the $1 -\alpha$ quantile of the the conditional likelihood ratio test. The $F$ distribution for the Anderson-Rubin test is based on the aforementioned assumption about Normal errors in model \eqref{eq:model} and our package \textbf{ivmodel} currently uses the $F$ distribution. However, one can also use the $\chi^2$ distribution as an asymptotic approximation if Normal errors are grossly unreasonable in data. As for the null distribution for the conditional likelihood ratio test and the associated quantile value $q_{1-\alpha}$, see \citet{andrews_performance_2007}.

\section{Dealing with invalid instruments}

\subsection{IV diagnostic}
\label{sec:diagnosis}

\citet{morgan_counterfactuals_2007} showed that assumption (A3) cannot
be completely verified. However, there is often concern that a putative
IV is invalid in applications. To assess the potential bias due to
non-exogeneity of the instruments, our \textbf{ivmodel} package
implements a graphical diagnosis of IV analysis proposed in
\citet{zhao_graphical_2018}. By assuming a single binary IV and the
control potential outcome depends linearly on only one covariate
$X_{ij}$,
\begin{equation}
  \label{eq:modelPotentialOutcome-1}
  \E(Y_i^{(0,0)} \mid Z_i, X_{i \cdot}) =  \kappa_j X_{ij},
\end{equation}
\citet{brookhart_preference_2007} derived the following bias formulas
for TSLS and OLS that do not adjust for any covariate.
\begin{align}
  \label{eq:bias-tsls}
  \text{bias}(\hat{\beta}_{\text{TSLS}}) &= \kappa_j \cdot \frac{\E[X_{ij}
    \mid Z_i = 1] - \E[X_{ij} \mid Z_i = 0]}{\E[D_i \mid Z_i = 1] -
    \E[D_i \mid Z_i = 0]},\\
  \label{eq:bias-ols}
  \text{bias}(\hat{\beta}_{\text{OLS}}) &= \kappa_j \cdot
                                          \big(\E[X_{ij} \mid D_i = 1] - \E[X_{ij} \mid D_i = 0] \big).
\end{align}
\citet{jackson2015toward} proposed to report a table of the
ratios between \eqref{eq:bias-tsls} and \eqref{eq:bias-ols} to assess
the potential advantage of an IV analysis over a standard regression
analysis. \citet{zhao_graphical_2018} further pointed out that a large
bias ratio might be misleading when the covariate is irrelevant
($\kappa_j \approx 0$) and suggested to use a diagnostic barplot to compare
\eqref{eq:bias-tsls} with \eqref{eq:bias-ols}.  Broadly speaking, if the bias from an IV analysis is smaller than the bias from a standard regression analysis (i.e. the ratio of biases is between $-1$ and $1$ or the difference between the two biases is large) and the aforementioned assumptions underlying the bias calculations are plausible, it suggests that an IV analysis is more helpful in reducing confounding bias than a standard regression analysis. In contrast, if the bias from an IV analysis is larger than the bias from a standard regression analysis (i.e. the ratio of biases is larger than $1$ or smaller than $-1$), a standard regression analysis may reduce more confounding than an IV analysis; see our data example in Section \ref{sec:diagnosis-1} for an example interpretation of confounding and bias reduction. When $Z$ or $D$ is not binary, we may replace the difference in conditional expectations in
\eqref{eq:bias-tsls} and \eqref{eq:bias-ols} by the corresponding OLS
slope coefficient.

We remark that the graphical diagnosis does not give a test the
validity of assumption (A3), as the simplifying assumption
\eqref{eq:modelPotentialOutcome-1} is different from
\eqref{eq:modelPotentialOutcome} that is used to define the residual
$\epsilon_i$. Furthermore, the bias formulas \eqref{eq:bias-tsls} and
\eqref{eq:bias-ols} only apply to the vanilla TSLS and OLS estimators
that do not adjust for any covariate. Thus, they do not equal the
true bias of the TSLS and OLS estimators that adjust for the
covariates $X_{i \cdot}$ (due to not controlling for other
  unmeasured confounders). Nevertheless, the diagnostic plot provides a way
to check if the IV is independent of any measured covariate and if not,
how much bias that dependence might incur. Alternatively, by
leveraging additional assumptions, some statistical tests have been
developed to falsify the validity of an IV (that is, to test
assumption (A3)); see \citet{glymour_credible_2012},
\citet{yang_dissonant_2014} and \citet{keele_falsification_2019}.

\subsection{Sensitivity analysis}
\label{sec:sensitivity}
Another way to deal with invalid instruments is through a sensitivity
analysis which examines the sensitivity of statistical tests for $H_0:
\beta = \beta_0$ to violations of (A3); see
\citet{diprete_assessing_2004}, \citet{small_sensitivity_2007},
\citet{kolesar_identification_2015} and \citet{conley_plausibly_2012}
for some examples. These papers all use test statistics which are
based on the TSLS estimator having an approximately normal
distribution, which breaks down in the presence of weak instruments
\citep{nelson_distribution_1990}. In this section, we explore a
sensitivity analysis based on the Anderson-Rubin test which is robust
to weak instruments and focus on the case where there is only one
instrument.

Formally, we revise the model in Section \ref{sec:model} to allow for an invalid instrument by adding another term $\delta\sigma(z'-z)$ to equation \eqref{eq:modelPotentialOutcome}.
 \begin{equation} \label{eq:sensmodelPotentialOutcome}
Y_{i}^{(d',z')} - Y_{i}^{(d,z)} = (d' - d) \beta +\delta\sigma(z'-z),\quad \E(Y_i^{(0,0,0)} \mid Z_{i \cdot}, X_{i \cdot}) =  X_{i \cdot}^T \kappa
\end{equation}
Here $\sigma$ is the standard variation of $\epsilon_i = Y_{i}^{(0,0,0)} - \E(Y_{i}^{(0,0,0)} \mid Z_{i \cdot}, X_{i \cdot})$ and serves as a scaling parameter. $\delta$ measures how much the instrument violates (A3) and lies within a range $\delta\in (\underline{\delta}, \bar{\delta})$ specified by the investigator. Then, the observed model for sensitivity analysis becomes:
\begin{equation} \label{eq:sensmodel}
Y_i = D_i \beta + X_{i \cdot}^T \kappa + \delta\sigma Z_{i \cdot} + \epsilon_i, \quad{} \E(\epsilon_i | Z_{i \cdot}, X_{i \cdot}) = 0, \quad{} \VAR(\epsilon_i | Z_{i \cdot}, X_{i \cdot}) = \sigma^2, \quad{} \delta\in (\underline{\delta}, \bar{\delta})
\end{equation}
If the error term has a normal distribution $\epsilon_i\sim N(0, \sigma^2)$, then hypothesis \eqref{eq:hyp} can be tested by using the AR test statistic $AR(\beta_0)$ in equation \eqref{eq:AR}. Under $H_0$, $AR(\beta_0)$ has a non-central $F$ distribution :
\begin{equation}
AR(\beta_0)\sim F_{1, n-p-1, \delta^2{Z^*}^TZ^*}
\end{equation}
Although $\delta$ is unknown and consequently we don't known exact null distribution of $AR(\beta_0)$ under $H_0$, we can look at the worst-case null distribution by setting $\delta$ to $\Delta=\max(|\underline{\delta}|, |\bar{\delta}|)$ and constructing a $1-\alpha$ sensitivity interval
\begin{equation}
\label{eq:sensCI}
CI_{1-\alpha}=\{\beta: AR(\beta_0)<F_{1, n-p-1, \Delta^2{Z^*}^TZ^*; 1-\alpha}\}
\end{equation}
More details about the above sensitivity analysis can be found in \citet{jiang_sensitivity_2017}.

\section{Power} \label{sec:power}
A power analysis concerns the probability of rejecting the null hypothesis $H_0: \beta=\beta_0$ when the true exposure effect is under the alternative $\beta-\beta_0=\lambda\neq0$. Often, power analysis is used to decide the number of samples to detect an effect size with certain probability. \citet{freeman_power_2013} presents a power formula for the TSLS estimator when used as a hypothesis test. \citet{jiang_sensitivity_2017} provides a power formula for the Anderson-Rubin test as well as a power formula for the sensitivity interval in Section \ref{sec:sensitivity}. In this section, we discuss these power formulas and the underlying assumptions that each make.

\citet{freeman_power_2013}'s power formula assumes only a single IV($L=1$) without any covariates $X$($p=0$); this setup is akin to model \eqref{eq:model} with $\kappa=0$. Under this setup, the TSLS estimator asymptotically follows a Normal distribution:
\begin{equation}
\label{eq:freeman}
\widehat{\beta}_{TSLS}\sim N\left(\beta, \quad{} \frac{\sigma^2}{n\cdot \VAR(D)\cdot \rho_{ZD}}\right)
\end{equation}
If the true exposure effect is $\beta-\beta_0=\lambda$, then the power of testing hypothesis \eqref{eq:hyp} is:
\begin{equation}
\label{eq:power_ttest}
\mbox{Power}=1+\Phi\brct{-z_{\alpha/2}-\frac{\lambda \rho_{ZD}\sqrt{n\cdot \VAR(D)}}{\sigma}}-\Phi\brct{z_{\alpha/2}-\frac{\lambda \rho_{ZD}\sqrt{n\cdot \VAR(D)}}{\sigma}}
\end{equation}
where $\alpha$ is the significance level (usually 0.05), $\Phi$ is the cumulative distribution function of the standard Normal distribution, $z_\alpha$ is the upper $\alpha$ quantile of the standard Normal distribution, i.e. $\Phi(-z_\alpha)=\alpha$, and $\rho_{ZD}$ is the correlation between $Z$ and $D$. 

The power formula for the Anderson-Rubin test is based on the original model \eqref{eq:model}, the first-stage model \eqref{eq:model_1st_stage}, and bivariate normality of the errors $(\epsilon_i, \eta_i)$, which we summarize below.
\begin{gather}
\begin{aligned}
Y^*&= D^* \beta +\epsilon^* \\
D^*&=Z^*\gamma+\eta^* \\
Y^*&=R_XY,~~D^*=R_XD,~~Z^*=R_XZ,~~\epsilon^*=R_X\epsilon,~~\eta^*=R_X\eta\\
(\epsilon, \eta)&\perp Z,~~~(\epsilon_i, \eta_i)^T\sim N\brct{\mathbf{0}, \Sigma},~~~
\Sigma=\bma{\sigma^2 & \rho\sigma\omega\\ \rho\sigma\omega & \omega^2},~~~\mbox{rank}(X)=p
\end{aligned}\label{eq:model_ARfull}
\end{gather}
If the true exposure effect is $\beta-\beta_0=\lambda$, the power of testing hypothesis \eqref{eq:hyp} using the Anderson-Rubin test is:
\begin{equation}
\label{eq:power_AR}
\mbox{Power}=1-\Psi_{1, n-p-L, \frac{(\gamma^T{Z^*}^TZ^*\gamma)\lambda^2}{\sigma^2+2\rho\sigma\omega\lambda+\omega^2\lambda^2}}(F_{1, n-p-L; 1-\alpha})
\end{equation}
where $F_{a, b; 1-\alpha}$ is the $1-\alpha$ quantile of the $F$ distribution with degrees of freedom $a$ and $b$. $\Psi_{a, b, k}(\cdot)$ is the cumulative distribution function of the non-central F distribution with degrees of freedom $a, b$ and non-central parameter $k$.

Finally, the power of the sensitivity analysis introduced in Section \ref{sec:sensitivity}
relies on model \eqref{eq:sensmodel}, the first stage model in \eqref{eq:model_1st_stage} and the bivariate Normality assumption of the errors $(\epsilon_i, \eta_i)$, which we summarize below
\begin{gather}
\begin{aligned}
Y^*&= D^* \beta +\delta\sigma Z^*+\epsilon^*\\
D^*&= Z^* \gamma +\eta^*\\
Y^*&=R_XY,~~D^*=R_XD,~~Z^*=R_XZ,~~\epsilon^*=R_X\epsilon,~~\eta^*=R_X\eta\\
(\epsilon, \eta)&\perp Z,~~~(\epsilon_i, \eta_i)^T\sim N\brct{\mathbf{0}, \Sigma},~~~
\Sigma=\bma{\sigma^2 & \rho\sigma\omega\\ \rho\sigma\omega & \omega^2},~~~\mbox{rank}(X)=p
\end{aligned}
\end{gather}
Suppose we are in the alternative where the true exposure effect is $\beta-\beta_0=\lambda$ and the instrument is valid ($\delta=0$). But, under the null hypothesis, we want to allow for the possibility that the instrument is invalid in the range $\delta \in (-\Delta,\Delta)$; this is referred to as the favorable situation in  \citet{rosenbaum_design_2010}. Then, the power of being able to reject the null hypothesis in favor of this favorable alternative for all $\delta \in (-\Delta,\Delta)$ is:
\begin{equation}
\label{eq:power_sen}
\mbox{Power}=1-\Psi_{1, n-p-1, \frac{\lambda^2\gamma^2{Z^*}^TZ^*}{\sigma^2+2\rho\sigma\omega\lambda+\omega^2\lambda^2}}(F_{1, n-p-1, \Delta^2{Z^*}^TZ^*; 1-\alpha})
\end{equation}
where $F_{a, b, c; 1-\alpha}$ is the $1-\alpha$ quantile of the non-central $F$ distribution with degrees of freedom $a, b$ and non-central parameter $c$. 
Generally speaking, when the instrument is weak and/or the sample size is small to moderate, the power formula for the TSLS test statistic may be biased and instead, \citet{jiang_sensitivity_2017} recommends using the AR test and its associated power formula \eqref{eq:power_AR}. Also, \citet{jiang_sensitivity_2017} shows that the AR test may have no power if $\Delta$ is large.

All three power formulas are implemented in \textbf{ivmodel}. \textbf{ivmodel} also provides functions to compute the minimum sample size needed to achieve a specific power at a specific $\beta$.

\section{Application} \label{sec:application}
In this section, we illustrate an application of \textbf{ivmodel} with the data set from \citet{card_using_1995}. The data is from the National Longitudinal Survey of Young Men (NLSYM), which has $n=3010$ individuals. Like \citet{card_using_1995}, we want to estimate the causal effect of education on log earnings by using a binary instrumental variable indicating whether the individual grew up in a place with a nearby 4-year college. The study also collected some exogenous variables for each study unit.

\subsection{Basic usage}
\label{sec:application-basic}
As discussed above, the outcome $Y$ is log earnings (\texttt{lwage}),
the exposure $D$ is (\texttt{educ}), the instrument $Z$ is
(\texttt{nearc4}). Other exogenous variables $X$ include subject's years of labor force
experience (\texttt{exper}) and its square (\texttt{expersq}), whether
the subject is black (\texttt{black}), whether the subject lived in
the South (\texttt{south}), and whether the subject is in a metropolitan area (\texttt{smsa}). While we are concerned that \texttt{exper} and \texttt{expsq} are endogeneous due them being derived variables from \texttt{educ} and \texttt{age}, surprisingly, \citet{card_using_1995}'s analysis treated \texttt{exper} as an exogenous variable (page 13 of \citet{card_using_1995}). He also found that treating \texttt{exper} as either endogenous or exogenous led to the same conclusions about education's return on earnings (Table 3 of \citet{card_using_1995}). More generally, treating experience as exogenous is common in labor economics; see \citet{heckman_earnings_2006} for a review. Overall, to focus on the software aspect of the paper, we recreate \citet{card_using_1995}, but alert the readers about this caveat.

We then use the function \emph{ivmodelFormula}, which takes in formulas of the style from \citet{zeileis_extended_2010} and that is also used in the package \textbf{AER}, and generate an \texttt{ivmodel} class object
\begin{verbatim}
R> cardfit = ivmodelFormula(lwage ~ educ + exper + expersq + black + south + smsa |
R+                 nearc4 + exper + expersq + black + south + smsa, data=card.data)
\end{verbatim}
\textbf{ivmodel} can also take non-formula environments as inputs by using the function \emph{ivmodel}.
\begin{verbatim}
R> Y = card.data[,"lwage"]
R> D = card.data[,"educ"]
R> Z = card.data[, "nearc4"]
R> Xname = c("exper", "expersq", "black", "south","smsa")
R> X = card.data[, Xname]
R> cardfit = ivmodel(Y=Y, D=D, Z=Z, X=X)
\end{verbatim}

After a \texttt{ivmodel} class object is generated, we can call \emph{summary} on the object to display all the relevant tests discussed above.

\begin{verbatim}
R> summary(cardfit)
\end{verbatim}
\begin{verbatim}
Call:
ivmodel(Y = Y, D = D, Z = Z, X = X)
sample size: 3010
_ _ _ _ _ _ _ _ _ _ _ _ _ _ _ _ _ _ _ _ _ _ _ _ _ _ _ _ _ _

First Stage Regression Result:

F=16.71759, df1=1, df2=3003, p-value is 4.4515e-05
R-squared=0.005536144,   Adjusted R-squared=0.005204987
Residual standard error: 1.942531 on 3004 degrees of freedom
_ _ _ _ _ _ _ _ _ _ _ _ _ _ _ _ _ _ _ _ _ _ _ _ _ _ _ _ _ _

Coefficients of k-Class Estimators:

              k Estimate Std. Error t value Pr(>|t|)
OLS    0.000000 0.074009   0.003505  21.113  < 2e-16 ***
Fuller 0.999667 0.128981   0.047601   2.710  0.00677 **
TSLS   1.000000 0.132289   0.049233   2.687  0.00725 **
LIML   1.000000 0.132289   0.049233   2.687  0.00725 **
---
Signif. codes:  0 `***' 0.001 `**' 0.01 `*' 0.05 `.' 0.1 ` ' 1
_ _ _ _ _ _ _ _ _ _ _ _ _ _ _ _ _ _ _ _ _ _ _ _ _ _ _ _ _ _

Alternative tests for the treatment effect under H_0: beta=0.

Anderson-Rubin test (under F distribution):
F=6.881108, df1=1, df2=3003, p-value is 0.0087552
95 percent confidence interval:
 [0.0383986007667666, 0.261183653633852]

Conditional Likelihood Ratio test (under Normal approximation):
Test Stat=6.881108, p-value is 0.0087552
95 percent confidence interval:
 [0.0383985832976054, 0.261183686557055]
\end{verbatim}

There are three main sections in the summary. The first section summarizes the first stage regression between the IV and the exposure. For example, in this data, the F statistic is 16.71759, which is greater than 10, indicating that the IV is not weak and TSLS estimator should be reasonable. The second section lists the results for several k-class estimators. The default $k$'s are $k=0$ (OLS), $k=1$ (TSLS), and $k$'s associated with LIML and Fuller. Here we only have one IV, so TSLS and LIML are the same. The estimated causal effect for TSLS is 0.132289, with a p-value around $7.25 * 10^{-3}$. This means that when increasing education by 1 year, ceteris paribus, log earnings will, on average, increase by $0.132289$. The last section provides AR and CLR confidence intervals, which are robust when weak instruments are present.

The function \emph{confint} calculates the confidence interval for various IV methods introduced above. Similarly, we also provide common functions such as \emph{coef}, \emph{fitted}, \emph{residuals}, \emph{vcov}, and \emph{model.matrix}. \emph{coef} extracts the coefficient of $\beta$. \emph{fitted} provides fitted values of $Y$. \emph{residuals} generates residuals $Y^* - D^* \widehat{\beta}_{k}$. \emph{vcov} computes the standard errors for each $\widehat{\beta}_{k}$. \emph{model.matrix}  extracts the design matrix used to fit the instrumental variables model.
\begin{verbatim}
R> confint(cardfit)
\end{verbatim}
\begin{verbatim}
             2.5%      97.5%
OLS    0.06713570 0.08088229
Fuller 0.03564754 0.22231476
TSLS   0.03575456 0.22882312
LIML   0.03575456 0.22882312
AR     0.03839860 0.26118365
CLR    0.03839858 0.26118369
\end{verbatim}

\subsection{Power and sample size}
Suppose the true causal effect of earnings is $\beta = 0.1$ and we want to compute the power of tests to reject the null hypothesis of no effect in favor of this alternative. \textbf{ivmodel} contains the function  \emph{IVpower}, which computes powers for the TSLS test statistic, the AR test, and the sensitivity analysis  test, with the default being the TSLS test statistic. In the example below, the power of the TSLS test statistic is 0.6864 and the power of the AR test is 0.8437.

\begin{figure}
  \centering\includegraphics[width=3 in]{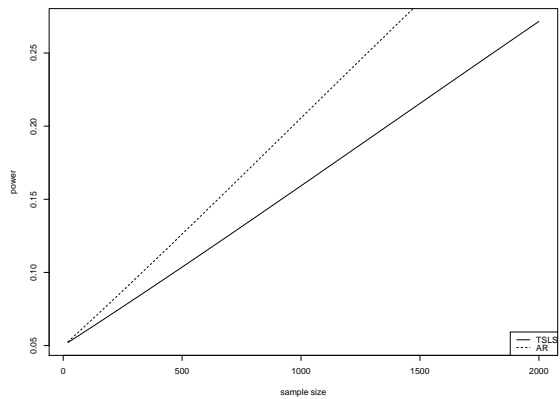}
  \caption{Power curves for the TSLS test statistic and the AR test under different sample sizes (x-axis). The alternative is fixed to be $\beta = 0.1$.}
  \label{fig:power}
\end{figure}

\begin{verbatim}
R> IVpower(cardfit,beta=0.1); IVpower(cardfit, type="AR",beta=0.1)
\end{verbatim}
\begin{verbatim}
[1] 0.5286761
[1] 0.5461072
\end{verbatim}
When there is only one instrument and the errors are Normally distributed with a known covariance matrix, the AR test is the uniformly most powerful test \citep{andrews_optimal_2006}, hence the large power. We can also compare the power under different sample sizes by plotting a figure of power as a function of sample size. Figure~\ref{fig:power} is a graphical output of power functions for the TSLS test statistic and the AR test as a function of sample size.

\begin{verbatim}
R> ngrid = (1:100)*20
R> plot(IVpower(cardfit, beta=0.1,n=ngrid)~ngrid,
        type="l", lty=1, ylab="power",xlab="sample size")
R> points(IVpower(cardfit, beta = 0.1,n=ngrid, type="AR")~ngrid,
        type="l", lty=2)
R> legend("bottomright", legend=c("TSLS", "AR"), lty=c(1, 2))
\end{verbatim}

Finally, \emph{IVsize} calculates the minimum sample size needed for achieving a certain power threshold. In the example below, we need a sample size of 3950 for the TSLS test statistic and 2679 for the AR test in order to reject the null in favor of the alternative $\beta = 0.1$ with 80\% probability.
\begin{verbatim}
R> IVsize(cardfit, beta=0.1,power=0.8)
R> IVsize(cardfit, beta=0.1,power=0.8, type="AR")
\end{verbatim}
\begin{verbatim}
[1] 5723
[1] 5482
\end{verbatim}

\subsection{Diagnostic}
\label{sec:diagnosis-1}

Often in an IV analysis, there is concern that the instrument may be
invalid. For example, in our dataset, there may be concern that
geographic or social features affect both the existence of a
nearby 4-year college and earnings of an individual, but not through
education. This issue can be seen from the diagnostic plot generated by
\emph{iv.diagnosis}.
\begin{verbatim}
R> output <- iv.diagnosis(Y = Y, D = D, Z = Z, X = X)
R> iv.diagnosis.plot(output)
\end{verbatim}

\begin{figure}[h!]
  \centering\includegraphics[width=3.5 in]{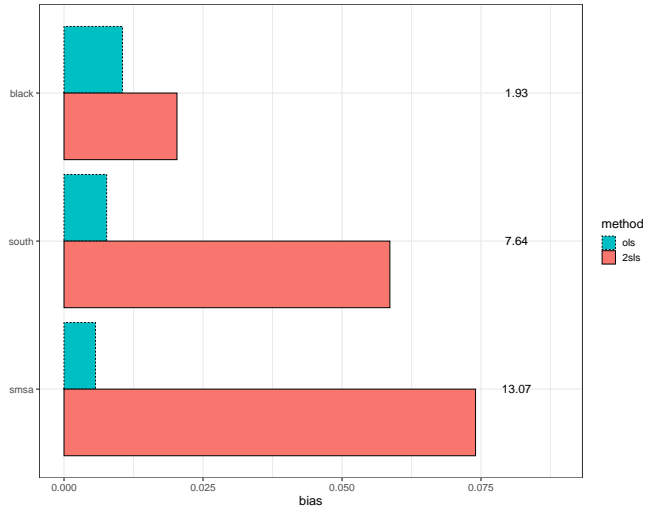}
  \caption{Diagnostic plot for the IV analysis. Each bar in each row represents the magnitude of the bias from not adjusting for a covariate. The number to the right represents the ratio of the biases.}
  \label{fig:diagnosis}
\end{figure}

The results are shown in Figure \ref{fig:diagnosis}. The red and blue
bars in Figure \ref{fig:diagnosis} are
estimated biases using \eqref{eq:bias-tsls} and \eqref{eq:bias-ols}
and the numbers on the right are the ratios between the two biases. The most striking
observation from Figure \ref{fig:diagnosis} is that the vanilla TSLS
estimator would have more than 13 times larger bias than the vanilla OLS
estimator if the control potential outcome depends linearly according
to \eqref{eq:modelPotentialOutcome-1} on \texttt{smsa}, and the
absolute bias would be as large as $0.07$. The bias ratios with respect to
\texttt{south} and \texttt{black} are also larger than $1$. We note
that these interpretations of Figure \ref{fig:diagnosis} depend on the
assumptions made in Section \ref{sec:diagnosis}. Specifically, the
simplifying assumption \eqref{eq:modelPotentialOutcome-1} that the
control potential outcome depends linearly on only one covariate is rather
strong, so the diagnostic plot should be interpreted with caveat in mind.

Nevertheless, the diagnostic plot indicates that, without controlling for any covariates, the
instrument---proximity to college---is correlated with geographic
features such as \texttt{south} and \texttt{smsa} that may also affect
the earnings. In particular, if we examine the correlation matrix of the
variables below, we see that the geographic features (\texttt{south} and
\texttt{smsa}) have much stronger correlation with both the instrument
and the outcome than labor force experience (\texttt{exper} and
\texttt{expersq}).
\begin{verbatim}
R> round(cor(cbind(Z, D, X, Y)), 2)
\end{verbatim}
\begin{verbatim}
            Z     D exper expersq black south  smsa     Y
Z        1.00  0.14 -0.06   -0.06 -0.08 -0.22  0.35  0.16
D        0.14  1.00 -0.65   -0.63 -0.27 -0.20  0.19  0.31
exper   -0.06 -0.65  1.00    0.97  0.14  0.11 -0.14  0.01
expersq -0.06 -0.63  0.97    1.00  0.13  0.12 -0.14 -0.02
black   -0.08 -0.27  0.14    0.13  1.00  0.34 -0.04 -0.30
south   -0.22 -0.20  0.11    0.12  0.34  1.00 -0.18 -0.28
smsa     0.35  0.19 -0.14   -0.14 -0.04 -0.18  1.00  0.23
Y        0.16  0.31  0.01   -0.02 -0.30 -0.28  0.23  1.00
\end{verbatim}
Overall, although we can use a TSLS estimator to adjust for these observed covariates like \texttt{south} and
  \texttt{smsa}, there may well be residual confounding that
  positively biases the IV analysis. This means the true causal effect
  of education on earning might not be as large as the estimate from
  TSLS.

  To further illustrate this point, Table \ref{tab:covariate-compare}
  compares the OLS and TSLS estimates obtained using \textit{ivmodel}
  when different covariates are adjusted for. When only adjusting for
  \texttt{exper}, \texttt{expersq}, \texttt{black} but not any
  geographic features, the TSLS estimate is 0.255. This estimate
  becomes closer to the OLS estimate as the geographic features are
  included, eventually dropping to 0.132. Both \texttt{south} and
  \texttt{smsa} are coarse measurements of the geography of survey
  participants. Had we obtained finer geographic features, the TSLS
  estimate might be even smaller.

  \begin{table}[h]
    \caption{A comparison of OLS and TSLS estimates adjusted for
      different sets of covariates.}
    \label{tab:covariate-compare}
  \begin{tabular}{l|rr|rr}
    \toprule
    & \multicolumn{2}{c|}{OLS} & \multicolumn{2}{c}{TSLS} \\
    Adjusted covariates & Estimate & Std.\ error & Estimate &
                                                                 Std.\
                                                                 error
    \\
    \midrule
    None & 0.052 & 0.003 & 0.188 & 0.026 \\
    \texttt{exper}, \texttt{expersq}, \texttt{black} & 0.082 & 0.004 &
                                                                       0.255
                                   & 0.038 \\
    \texttt{exper}, \texttt{expersq}, \texttt{black}, \texttt{south} &
                                                                       0.078
                               & 0.004 & 0.221 & 0.041 \\
    \texttt{exper}, \texttt{expersq}, \texttt{black}, \texttt{smsa} &
                                                                       0.076
                               & 0.004 & 0.177 & 0.046 \\
    \texttt{exper}, \texttt{expersq}, \texttt{black}, \texttt{south}, \texttt{smsa} &
                                                                       0.074
                               & 0.004 & 0.132 & 0.049 \\
    \bottomrule
  \end{tabular}
  \end{table}

\subsection{Sensitivity analysis}

We can also perform a sensitivity analysis to assess the sensitivity
of our analysis to invalid IV. The user needs to specify the likely
range of departure from assumption (A3), captured by the parameter
$\delta$ in \eqref{eq:sensmodel}. Roughly speaking, the parameter $\delta
\sigma$ captures how much a unit change in the invalid instrument
\texttt{near4c} will change the outcome \texttt{lwage} the regression
model \eqref{eq:sensmodel}, either through a direct causal effect of
\texttt{near4c} on \texttt{lwage} or through correlation of
\texttt{near4c} with unincluded determinants of \texttt{lwage} like
\texttt{south}.

One way to gauge how large $\delta$
might be is to first fit a standard regression model for the outcome conditional on the education and exogenous covariates.
\begin{verbatim}
R> summary(lm(lwage ~ educ + exper + expersq + black + south + smsa,
              data = card.data))
\end{verbatim}
\begin{verbatim}
Call:
lm(formula = lwage ~ educ + exper + expersq + black + south +
    smsa, data = card.data)

Residuals:
     Min       1Q   Median       3Q      Max
-1.59297 -0.22315  0.01893  0.24223  1.33190

Coefficients:
              Estimate Std. Error t value Pr(>|t|)
(Intercept)  4.7336643  0.0676026  70.022  < 2e-16 ***
educ         0.0740090  0.0035054  21.113  < 2e-16 ***
exper        0.0835958  0.0066478  12.575  < 2e-16 ***
expersq     -0.0022409  0.0003178  -7.050 2.21e-12 ***
black       -0.1896315  0.0176266 -10.758  < 2e-16 ***
south       -0.1248615  0.0151182  -8.259  < 2e-16 ***
smsa         0.1614230  0.0155733  10.365  < 2e-16 ***
---
Signif. codes:  0 `***' 0.001 `**' 0.01 `*' 0.05 `.' 0.1 ` ' 1

Residual standard error: 0.3742 on 3003 degrees of freedom
Multiple R-squared:  0.2905,    Adjusted R-squared:  0.2891
F-statistic: 204.9 on 6 and 3003 DF,  p-value: < 2.2e-16
\end{verbatim}

Imagine a unmeasured confounder $U$ similar to \emph{south}, in the sense
that $U$ has the same effect on the instrument and the outcome as
\emph{south}. Further, suppose $U$ is independent of the other measured
covariates (this is slightly different from \emph{south} which is
weakly correlated with the other covariates). Then we expect the
$\delta$ corresponding to such $U$ is about $0.22$ (correlation of \texttt{south} with
  \texttt{nearc4}) $\times~0.12$ (coefficient of \texttt{south} in the
  regression for \texttt{lwage}) $/~0.3742$ (estimated $\sigma$ in the
  regression for \texttt{lwage}) $\approx 0.07$. Thus, we might assume
  the range for the sensitivity parameter is $\delta\in (-0.07, 0.07)$
  to reflect having a covariate like \texttt{south}.

 To perform a sensitivity
analysis, we can call the function \emph{ivmodel} specifying the range
of the sensitivity parameter.
  \begin{verbatim}
 R> cardfit.sens =ivmodel(Y=Y, D=D, Z=Z, X=X, deltarange=c(-0.07, 0.07))
 R> summary(cardfit.sens)
  \end{verbatim}
  \begin{verbatim}
Anderson-Rubin test:
Sensitivity analysis with deltarange [ -0.07 ,  0.07 ]:
non-central F=6.881108, df1=1, df2=3003, ncp=2.71656, p-value is 0.16499
95 percent confidence interval:
 [ -0.0538384077784691 , 0.53548242970625 ]
  \end{verbatim}
  We see that if there is an unmeasured confounder $U$ that exhibits similar
  behavior as the variable \texttt{south}, we would retain the null hypothesis of no effect
  when we use the Anderson-Rubin test statistic. The p-value from the sensitivity
analysis is about $0.16$, suggesting that education does not have aa
significant positive effect towards earnings if the instrument
is invalid due to an unmeasured confounder $U$ with $\Delta$ around 0.07.

We also performed a ``synthetic'' sensitivity
analysis where we intentionally drop the variable \texttt{south} and see if the sensitivity
interval in the IV model without \texttt{south} matches the confidence interval in the IV model with \texttt{south}.
\begin{verbatim}
R> XwoSouth = X[,c("exper", "expersq", "black","smsa")]
R> cardfit2=ivmodel(Y=Y, D=D, Z=Z, X=XwoSouth, deltarange=c(-0.07, 0.07))
R> summary(cardfit2)
\end{verbatim}
\begin{verbatim}
Anderson-Rubin test:
Sensitivity analysis with deltarange [ -0.07 ,  0.07 ]:
non-central F=16.05672, df1=1, df2=3004, ncp=2.785717, p-value is 0.0097825
95 percent confidence interval:
 [ 0.0379720391935471 , 0.513984691572249 ]
\end{verbatim}
Notice that the lower end of this sensitivity interval is nearly identical to
the confidence interval for the Anderson-Rubin test that used all the
covariates in Section \ref{sec:application-basic}.

Finally, we can compute the power to
detect the favorable alternative under the null
hypothesis of no effect, but with a potentially invalid IV. For example,
suppose the true effect is $\beta^* = 0.25$. Then, the power to reject
the null of no effect in favor of this alternative with a $\delta \in (-0.07,0.07)$ is
22.7\% and we need at least 23,230 samples to increase this power to 80\%.
\begin{verbatim}
R> IVpower(cardfit.sens, beta=0.25, type="ARsens")
R> IVsize(cardfit3, beta=0.25, power=0.8, type="ARsens")
\end{verbatim}
\begin{verbatim}
[1] 0.2265288
[1] 23230
\end{verbatim}

\section{Summary}
The package \pkg{ivmodel} provides a unified implementation of instrumental variables methods in the case of one endogenous variable. The package contains a general class of estimators, $k$-class estimators, to estimate the parameter $\beta$. The package also contains methods that can deal with violations of instrumental variables assumptions, (A2) and (A3). First, for violations of (A2), the package contains two confidence intervals that are fully robust to weak instruments. For (A3), the package contains methods for sensitivity analysis for the range of violation. Additionally, the package contains power formulas to guide designs of future instrumental variables studies. As our data example in Section \ref{sec:application} demonstrated, our package provides an easy and unified way of conducting a comprehensive instrumental variables analysis with a given data by providing both ways to estimate the parameter of interests, ways to assess the sensitivity of our estimates to violations of IV assumptions, and ways to plan for future IV studies  in the form of a power analysis.


\acks{The research of Hyunseung Kang was supported in part by NSF Grant DMS 1811414.}


\newpage

\appendix




\vskip 0.2in
\bibliography{mainbib_2}

\end{document}